\documentstyle[amssymb,eqsecnum,aps,epsf]{revtex}

\begin{document}

\twocolumn[\hsize\textwidth\columnwidth\hsize\csname@twocolumnfalse\endcsname

\title{{\bf Cooper pair dispersion relation for weak to strong coupling}}
\author{Sadhan K. Adhikari$^1$, M. Casas$^2$, A. Puente$^2$, A. Rigo$^2$, 
M. Fortes$^3$, M. A. Sol\'{\i}s$^3$,\\ M. de Llano$^4$, Ariel A.
Valladares$^4 $, O. Rojo$^5$}
\address{$^1$Instituto de F\'{\i}sica Te\'{o}rica, Universidade Estadual
Paulista, 
01405-900 S\~{a}o Paulo, SP, Brazil}
\address{$^2$Departament de F\'{\i}sica, Universitat de les Illes
Balears, 
07071 Palma de Mallorca, Spain}
\address{$^3$Instituto de F\'{\i}sica, Universidad Nacional Aut\'{o}noma
de
M\'{e}xico,
01000 M\'{e}xico, DF, Mexico}
\address{$^4$Instituto de Investigaciones en Materiales, Universidad
Nacional
Aut\'{o}noma de M\'{e}xico,
04510 M\'{e}xico, DF, Mexico}
\address{$^5$PESTIC, Secretar\'{\i}a Acad\'{e}mica \& CINVESTAV, IPN,
04430 
M\'{e}xico DF, Mexico }
\maketitle

\begin{abstract}
Cooper pairing in two dimensions is analyzed with a set of renormalized
equations to determine its binding energy for any fermion number density and
all coupling assuming a generic  pairwise residual
interfermion interaction. \ Also considered are Cooper pairs (CPs) with
nonzero center-of-mass momentum (CMM)---usually neglected in BCS
theory---and their binding energy  is expanded analytically in powers
of the CMM up to quadratic terms. \ A Fermi-sea-dependent {\it linear} term
in the CMM dominates the pair excitation energy in weak coupling (also called the BCS
regime) while the more familiar quadratic term prevails in strong coupling
(the Bose regime). The crossover, though strictly unrelated to BCS theory {\it 
per se,} is studied numerically as it is expected to play a central role in
a model of superconductivity as a Bose-Einstein condensation of CPs where
the transition temperature vanishes for all dimensionality $d\leq 2$ for
quadratic dispersion, but is {\it nonzero }  for all $d\geq
1$ for linear dispersion.

PACS \#: 74.20.Fg; 64.90+b; 05.30.Fk; 05.30.Jp
\end{abstract}

\vskip1.5pc]

\newpage

The original Cooper pair (CP) problem \cite{coop} in two (2D) and three (3D)
dimensions possesses ultraviolet divergences in momentum space that are
usually removed via interactions regularized with large-momentum cutoffs 
\cite{adhik}. One such regularized potential is the BCS {\it model
interaction }which is of great practical use in studying Cooper pairing \cite
{coop} and superconductivity \cite{bcs}. Although there are controversies
over the precise pairing mechanism, and thus over the microscopic
Hamiltonian appropriate for high-$T_{c}$ superconductors, some of the
properties of these materials have been explained satisfactorily within a
BCS-Bose crossover picture \cite{mr,adh1,adhi,1964} via a renormalized BCS
theory for a short-range interaction. In the weak-coupling limit of the
BCS-Bose crossover description one recovers the pure mean-field BCS theory
of weakly-bound, severely-overlapping CPs. \ For strong coupling (and/or low
density) well-separated, nonoverlapping (so-called ``local'') pairs appear 
\cite{mr} in what is known as the Bose regime. It is of interest to detail
how renormalized Cooper pairing itself evolves {\it independently} of the
BCS-Bose crossover picture in order to then discuss the possible
Bose-Einstein (BE) condensation (BEC) of such pairs. We address this here in
a {\it single}-CP {\it picture}, while considering also the important case
(generally neglected in BCS theory) of non-zero center-of-mass-momentum
(CMM) CPs that are expected to play a significant role in BE condensates at
higher temperatures. \ 

In this Report we derive a renormalized Cooper equation for a pair of
fermions interacting via either a zero- or a finite-range interaction. We
find an analytic expression for the CP excitation energy up to terms
quadratic in the CMM which is valid for any coupling. For weak coupling only
the {\it linear} term dominates, as it also does for the BCS model
interaction \cite{physc}. \ The linear term was mentioned for 3D as far back
as 1964 (Ref. \cite{sch64} p. 33). For strong coupling we now find that the 
{\it quadratic }term dominates and is just the kinetic energy of the
strongly-bound composite pair moving in vacuum.

The CP dispersion relation enters into each summand in the BE distribution
function of the boson number equation from which the critical temperature $%
T_{c}$ of BEC of CPs is extracted. The linear dependence on the CMM of CP
binding for weak coupling leads to novel transition properties even in a
heuristic BEC picture of superconductivity \cite{pla} as BE-condensing CPs.
It is well-known that BEC is possible only for dimensionalities $d>2$ for
usual nonrelativistic bosons with quadratic dispersion; this limitation
reappears in virtually all BEC schemes thus far applied to describe
superconductivity (Refs. \cite{18a,Blatt} among others). But for bosons with
a {\it linear} dispersion relation as found here in weak coupling, BEC can
now occur for all $d>1$. \ Here we discuss the CP dispersion relation only
in 2D. We have also performed a similar analysis in 3D and obtained the
linear to quadratic crossover in the dispersion relation.

Consider a two-fermion system in the CM frame, with each fermion of mass $m$
, interacting via the purely attractive short-range separable potential \cite
{18a} 
\begin{equation}
V_{pq}=-v_{0}g_{p}g_{q},  \label{1}
\end{equation}
where $v_{0}\geq 0$ is the interaction strength and the $g_{p}$'s are the
dimensionless form factors $g_{_{p}}\equiv (1+p^{2}/p_{0}^{2})^{-1/2}$,
where the parameter $p_{0}$ is the inverse range of the potential so that,
e.g.,$\ p_{0}\rightarrow \infty $ implies $g_{_{p}}=1$ and corresponds to
the contact or delta potential $V(r)=-v_{0}\delta ({\bf r})$. The
interaction model (\ref{1})\ mimics a wide variety of possible dynamical
mechanisms in superconductors: mediated by phonons, or plasmons, or
excitons, or magnons, etc., or even a purely electronic interaction. In the
first instance mentioned, two terms of the form (1) can simulate  a
coulombic interfermion repulsion surrounded by a longer-ranged
electron-phonon attraction. The momentum-space Schr\"{o}dinger eigenvalue
equation for a two-particle bound state in vacuum with binding energy $%
B_{2}\geq 0$ for interaction (\ref{1}) is \cite{adh1} 
\begin{equation}
\frac{1}{v_{0}}=\sum_{k}\frac{g_{k}^{2}}{B_{2}+\hbar ^{2}k^{2}/m},\quad
\;\;\;\text{ }  \label{4}
\end{equation}
where $k$ is the wavenumber in the CM frame and $\hbar ^{2}k^{2}/2m$ the
single-particle energy.

On the other hand, the CP equation for two fermions above the Fermi surface
with momenta wavevectors ${\bf k}_{1}$ and ${\bf k}_{2}$ (and arbitrary CMM
wavevector ${\bf K}\equiv {\bf k}_{1}+{\bf k}_{2}$) is given by 
\begin{equation}
\biggr[{\frac{\hbar ^{2}k^{2}}{m}-E_{K}+\frac{\hbar ^{2}K^{2}}{4m}}\biggr] %
C_{k}=-\sum_{q}{}^{^{\prime }}{}V_{kq}C_{q},  \label{6}
\end{equation}
where ${\bf k}\equiv \frac{1}{2}({\bf k}_{1}-{\bf k}_{2})$ is the relative
momentum, $E_{K}\equiv 2E_{F}-\Delta _{K}$ the total pair energy, $\Delta
_{K}\geq 0$ the CP binding energy, $C_{q}\equiv \langle q\mid \Psi \rangle $
its wave function in momentum space, and the prime on the summation implies
restriction to states {\it above} the Fermi surface: viz., $|{\bf k}\pm {\bf %
K}/2|>k_{F}$. For the separable interaction (\ref{1}) Eq. (\ref{6}) becomes 
\begin{equation}
\sum_{k}{}^{^{\prime }}{}\frac{g_{k}^{2}}{\hbar ^{2}k^{2}/m+\Delta
_{K}-2E_{F}+\hbar ^{2}K^{2}/4m}=\frac{1}{v_{0}}.  \label{7}
\end{equation}
Although the summand in Eq. (\ref{7}) is angle-independent, the restriction
on the sum arising from the filled Fermi sea is a function of the relative
wave vector ${\bf k}$, and therefore angle-{\it dependent}. The potential
strength $v_{0}$ can be eliminated between Eqs. (\ref{4}) and (\ref{7})
leading to the renormalized CP equation 
\begin{eqnarray}
&\sum_{k}&\frac{g_{k}^{2}}{B_{2}+\hbar ^{2}k^{2}/m}  \nonumber \\
&=&\sum_{k}{}^{^{\prime }}\frac{g_{k}^{2}}{\hbar ^{2}k^{2}/m+\Delta
_{K}-2E_{F}+\hbar ^{2}K^{2}/4m},  \label{8}
\end{eqnarray}
Instead of the arbitrary cutoff usually employed in dealing with delta
interactions, in Eq. (\ref{8}) we rely on physical ``observables'' for the
sake of renormalization, viz., the ground-state binding energy $B_{2}$ in
vacuum. \ The sums in Eq. (\ref{8}) can be transformed to integrals; the
restriction in the second term arising from the filled Fermi sea leads to
two different expressions depending on whether $\widetilde{K}\equiv K/k_{F}$
is $<2$ or $>2$, as discussed in the Appendix. \ Letting all variables be
dimensionless by expressing them either in units of the Fermi wavenumber $%
k_{F}$ or of the Fermi energy $E_{F}\equiv \hbar ^{2}k_{F}^{2}/2m$, viz., $%
\xi \equiv k/k_{F}$, $\widetilde{B}_{2}\equiv B_{2}/E_{F}$, $\widetilde{
\Delta }_{K}\equiv \Delta _{K}/E_{F}$, etc., we define $\alpha
_{K}^{2}\equiv 1-\widetilde{\Delta }_{K}/2-\widetilde{K}^{2}/4\equiv -\beta
_{K}^{2}$, and $\theta $ the angle between wavevectors ${\bf k}$ and ${\bf K}
$ so that $\xi _{0}(\theta )\equiv \sqrt{1-\widetilde{K}^{2}\sin ^{2}\theta
/4}+\widetilde{K}\cos \theta /2$ and $\xi _{0}^{\prime }(\theta )\equiv - 
\sqrt{1-\widetilde{K}^{2}\sin ^{2}\theta /4}+\widetilde{K}\cos \theta /2$. \
For a zero-range interaction, $g_{k}=1$, after some algebra one gets 
\begin{equation}
\int_{0}^{\pi /2}d\theta \ln [\xi _{0}^{2}(\theta )-\alpha _{K}^{2}]=\frac{
\pi }{2}\ln {\frac{\widetilde{B}_{2}}{2}},\quad \widetilde{K}<2,  \label{1a}
\end{equation}
\begin{equation}
\frac{\pi }{2}\ln [\beta _{K}^{2}]-\int_{0}^{\theta _{0}}d\theta \ln \frac{
\xi _{0}^{\prime 2}(\theta )+\beta _{K}^{2}}{\xi _{0}^{2}(\theta )+\beta
_{K}^{2}}=\frac{\pi }{2}\ln {\frac{\widetilde{B}_{2}}{2}},\quad \;\widetilde{
K}>2,  \label{1b}
\end{equation}
where $\theta _{0}=\arcsin (2/\widetilde{K})<\pi /2$. \ For $\widetilde{K}=0$
only Eq. (\ref{1a}) applies, in which case $\xi _{0}(\theta )=1,$ $\alpha
_{K}^{2}\equiv \alpha _{0}^{2}=1-\widetilde{\Delta }_{0}/2$ and we obtain
the surprising result $\widetilde{\Delta }_{0}=\widetilde{B}_{2}$, i.e., for
an attractive delta interaction the vacuum and CP binding energies for zero
CMM coincide for {\it all} coupling, a result apparently first obtained in
Ref. \cite{adh1}. For a nonzero CMM the CP binding energies $\Delta _{K}$
can be calculated from Eqs. (\ref{1a}) and ( \ref{1b}). For $\widetilde{K}
\neq 0$ a minimum threshold value of $B_{2}/E_{F}$ is found to be required
to bind a CP.

Equations (\ref{1a}) and (\ref{1b}) can be solved numerically for the CP
binding $\Delta _{K}$ for any CMM. \ For small CMM only Eq. (\ref{1a}) is
relevant; and this equation for small but non-zero $\widetilde{K}$ and for $%
\widetilde{K}=0$ can be subtracted one from the other. This gives the
small-CMM expansion {\it valid for any coupling }$B_{2}/E_{F}\geq 0$, 
\begin{eqnarray}
\varepsilon _{K} &\equiv &({\Delta _{0}-\Delta _{K})}=\frac{2}{\pi }\hbar
v_{F}K  \nonumber \\
&+&\left[ 1-\left\{ 2-\left( \frac{4}{\pi }\right) ^{2}\right\} \frac{E_{F}}{
B_{2}}\right] \frac{\hbar ^{2}K^{2}}{2(2m)}+O(K^{3}),  \label{dk2}
\end{eqnarray}
where a nonnegative {\it CP excitation energy }$\varepsilon _{K}$ has been
defined, and the Fermi velocity $v_{F}$ comes from $E_{F}/k_{F}=\hbar
v_{F}/2 $. The leading term in (\ref{dk2}) is linear in the CMM, followed by
a quadratic term. The linear CP dispersion term should not be confused with
that of the many-body (collective) excitation spectrum in weak coupling. \
Only CPs can undergo BEC while bosonic ``excitations'' (or modes or phonons)
cannot since the former are {\it fixed }in number while the latter are not.
\ Indeed, the particle-hole [sometimes called the Anderson-Bogoliubov-Higgs
(ABH)] modes of excitation energy $\hbar v_{F}K/\sqrt{d}$ \cite{bel} in $d$
dimensions in the zero coupling limit are {\it larger} than the
weak-coupling CP dispersion energies $(2/\pi )\hbar v_{F}K$ and ${\frac{1}{2}
}\hbar v_{F}K$ (Ref. \cite{sch64} p. 33) in 2D and 3D \cite{sch64},
respectively, while in 1D they happen to coincide---in spite of the fact
that CPs and ABH-like modes are physically distinct entities.\ The
coefficient of the quadratic term in (\ref{dk2}) changes sign at $%
B_{2}/E_{F}={\Delta _{0}/E}_{F}\simeq 0.379,$ as one goes from weak ($B_{2}={%
\ \ \Delta _{0}}\ll E_{F}$) to strong ($B_{2}={\Delta _{0}}\gg E_{F}$)
coupling. \ If $v_{F}$ (or $E_{F}$) $\rightarrow 0$ explicitly (dilute
limit) the first two terms of Eq. (\ref{dk2}) reduce simply to 
\begin{equation}
\varepsilon _{K}\rightarrow \frac{\hbar ^{2}K^{2}}{2(2m)}\text{, }
\label{scl}
\end{equation}
for {\it any} coupling. \ This is clearly just the familiar nonrelativistic
kinetic energy in vacuum of the composite (so-called ``local'') pair of mass 
$2m$ and CMM $K$. The same result (\ref{scl}) is also found to hold in 3D.





Figure 1 shows exact numerical results for the zero-range potential ($%
g_{k}=1 $) for different couplings of a CP excitation energy $\varepsilon {%
_{K}/\Delta }_{0}$ as function of CMM $K/k_{F}$, both dimensionless. We note
that the CPs {\it break up} whenever $\Delta _{K\text{ }}$ turns from
positive to negative, i.e., vanishes, or by Eq. (\ref{dk2}) when $%
\varepsilon {\ _{K}/\Delta }_{0}{=1}$. These points are marked in the figure
by dots. In addition to the exact results obtained by solving Eqs. (\ref{1a}%
) and (\ref{1b}), we also exhibit the results for the linear approximation
[first term on the right-hand side of Eq. (\ref{dk2}), dot-dashed lines,
virtually coinciding with the exact curve for all $B_{2}/E_{F}\lesssim 0.1$]
as well as for the quadratic approximation (dashed parabolas) as given by
Eq. (\ref{scl}) for stronger couplings. For weak enough coupling or large
enough fermion density at any nonzero coupling the exact dispersion relation
is virtually linear---{\it in spite of the divergence} of the isolated
quadratic term in Eq. (\ref{dk2}) as $B_{2}/E_{F}\rightarrow 0$. \ As
coupling is increased the quadratic dispersion relation (\ref{scl}) slowly
begins to dominate. The crossover from a linear to a quadratic dispersion
relation manifests itself by a change in curvature from concave down to
concave up---these two regions being separated by an inflection point that
moves down towards the origin as coupling is increased to infinity.



Figure 2 shows the CP excitation energy $\varepsilon {_{K}/\Delta }_{0}$ as
a function of CMM $K/k_{F}$ calculated for the finite-range interaction form
factor $g_{p}=(1+p^{2}/p_{0}^{2})^{-1/2}$ with $p_{0}=k_{F}$ for weak to
moderate coupling. To compare, we also plot the zero-range result as well as
the linear relation given by the first term on the right-hand side of Eq. (%
\ref{dk2}). The finite-range curves are closer to the corresponding
zero-range ones if labeled by $\Delta _{0}/E_{F}$ instead of by $B_{2}/E_{F}$
, as was done with all four sets of curves.

Figure 3 exhibits $\varepsilon {_{K}/\Delta }_{0}$ as a function of $K/k_{F}$
for the finite-range interaction with $p_{0}=k_{F}$ for stronger couplings.
In this case there is no special advantage in labeling the dispersion curves
by $\Delta _{0}$ so $B_{2}$ was used with results for $B_{2}/E_{F}=$ 3, 10
and 20 shown. \ In the zero-range case the curves gradually tend to the
quadratic form as $B_{2}$ increases. For finite-range, $p_{0}=k_{F}$, the
curves develop a maximum followed by a minimum with a point of inflection in
between. \ The slope at the point of inflection is now negative. Although
each curve tends to a quadratic form for large enough $K/k_{F}$, they are
quite different from it for small $K/k_{F}$. These ``looped'' dispersion
curves are reminiscent of the ``roton'' excitation spectrum \cite{Landau} in
liquid $^{\text{4}}$He.

To summarize, the single CP problem with non-zero CMM is tracked as it
evolves in varying the interfermion short-range pair interaction from weak
to strong or varying fermion density from high to low, respectively, for any
fixed nonzero coupling. The CP excitation energy is exhibited as a function
of its CMM. For weak coupling, the excitation energy is a {\it linear}
dispersion relation in the CMM, and changes very gradually to a {\it %
quadratic } relation as coupling increases. For a zero-range pair
interaction in the dispersion curve one typically has a point of inflection
with a positive slope separating a region of concave-down curvature for
small CMM from a region of concave-up curvature for large CMM. For
finite-range pair interactions of sufficiently long range the slope at the
point of inflection changes from positive to zero and eventually becomes
negative. This leads to maxima and ``roton-like'' minima in the CP\
dispersion curves. These results will play a critical role in a model of
superconductivity based on BE condensation of CPs as they will yield, {\it %
even in 2D }as in the cuprates {\it ,} BEC transition temperatures $T_{c}$
that interpolate between nonzero values in weak coupling with a linear CP
dispersion relation down to the expected $T_{c}\equiv 0$ value in strong
coupling with a quadratic relation.

\bigskip

{\bf Appendix:} The restriction that both particles lie above the Fermi sea
in Eq. (\ref{8}) can be written as 
\begin{equation}
\left( {\bf k}/k_{F}\pm {\bf K}/2k_{F}\right) ^{2}-1=\xi ^{2}\pm \xi 
\widetilde{K}\cos \theta +\widetilde{K}^{2}/4-1\geq 0,  \label{a1}
\end{equation}
where $\xi \;{\bf \equiv \;}k/k_{F}$ and $\widetilde{K}\equiv K/k_{F}$. The
equality leads to two pairs of roots in $\xi $, say $\xi _{1,2}=-a\pm b$ and 
$\xi _{3,4}=a\pm b$, where $a\equiv (\widetilde{K}/2)\cos \theta $, $b\equiv 
\sqrt{1-(\widetilde{K}^{2}/4)\sin ^{2}\theta }$, and $\theta $ the angle
between ${\bf k}$ and ${\bf K}$.

For $\widetilde{K}<2$, $b>a$, one root of the two pairs is positive and the
other negative. Thus, Eq. (\ref{a1}) can be satisfied provided that $\xi $ $%
>\xi _{1}, \xi _{2}, \xi _{3}, \xi _{4}$, or specifically, if $\xi >\xi
_{0}(\theta )\equiv a+b. $

For $\widetilde{K}>2$ and $\theta >\theta _{0}\equiv \arcsin (2/\widetilde{K}
)$, $b$ becomes imaginary and Eq. (\ref{a1}) is satisfied for all $\xi $.
Therefore, there is no restriction in the integration over $\xi $. However,
for $\widetilde{K}>2$ and $\theta <\theta _{0}$, $b<a$ the pair of roots $%
\xi _{1,2}$ are both negative while the pair $\xi _{3,4}$ are both positive
(with $\xi _{3}>\xi _{4}$). Consequently, in both cases Eq. (\ref{a1}) is
satisfied only if $\xi $ is in the interval $[0,\xi _{0}^{\prime }(\theta
)\equiv a-b],$ and in the interval $\left[ \xi _{0}(\theta ),\infty \right] $
, respectively. Using these restrictions on the $\xi $ integration in Eq. ( 
\ref{8}) one eventually arrives at Eqs. (\ref{1a}) and (\ref{1b}).

\smallskip

{\large {\bf Acknowledgments:}} M.deLl. thanks S. Fujita for discussions,
D.M. Eagles for reading the manuscript, and V.V. Tolmachev for extensive
correspondence. Partial support from UNAM-DGAPA-PAPIIT (M\'{e}xico) \#
IN102198, CONACyT (M\'{e}xico) \# 27828 E, DGES (Spain) \# PB95-0492 and
FAPESP\ (Brazil) is gratefully acknowledged.

{\ {\bf Fig. 1.} Dimensionless CP excitation energy $\varepsilon _{K}/\Delta
_{0}$ {\it vs} $K/k_{F\text{ }}$, calculated from Eqs. (\ref{1a}) and (\ref
{1b}) for different couplings $B_{2}/E_{F}$, full curves. \ The dot-dashed
line is the linear approximation (virtually coincident with the exact curve
for $B_{2}/E_{F}$ $\lesssim $ $0.1$) while the dashed curve is the quadratic
term of Eq. (\ref{scl}). Dots denote values of CMM wavenumber where the CP\
breaks up, i.e., where $\Delta _{K}$ $\equiv 0$. \ }

{\ {\bf Fig. 2.} Same as Fig. 1 but for couplings expressed as $\Delta
_{0}/E_{F}$. The dot-dashed line is the linear approximation; the dashed
curve is the result for the finite-range interaction $p_{0}=k_{F}$, and the
full curve is the zero-range result. For the finite-range potential $\Delta
_{0}/E_{F}=0.1,$ $0.5,$ $1.0$ and $2.0$ correspond to $B_{2}/E_{F}=0.469,$ $%
1.4,$ $2.45$ and $4 $, respectively. Dots and squares mark values of CMM
wavenumber where the CP\ breaks up.}

{\ {\bf Fig. 3.} Same as Fig. 1 but including also finite range at stronger
couplings $B_{2}/E_{F}$. The full curve is the exact zero-range result; the
short-dashed one the quadratic approximation; the long-dashed one the exact
finite-range result with $p_{0}=k_{F}$. Each set of three curves is labeled
by different values of $B_{2}/E_{F}$.}

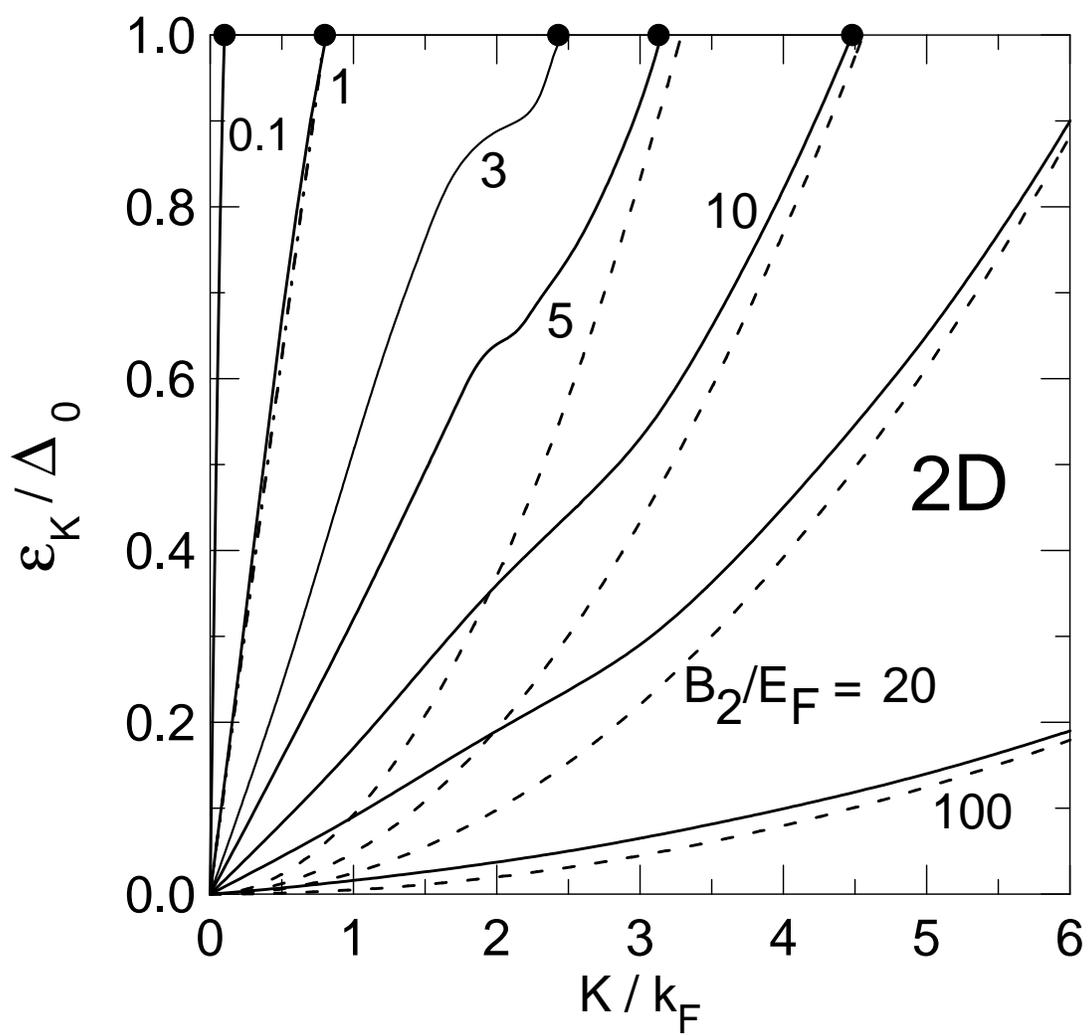
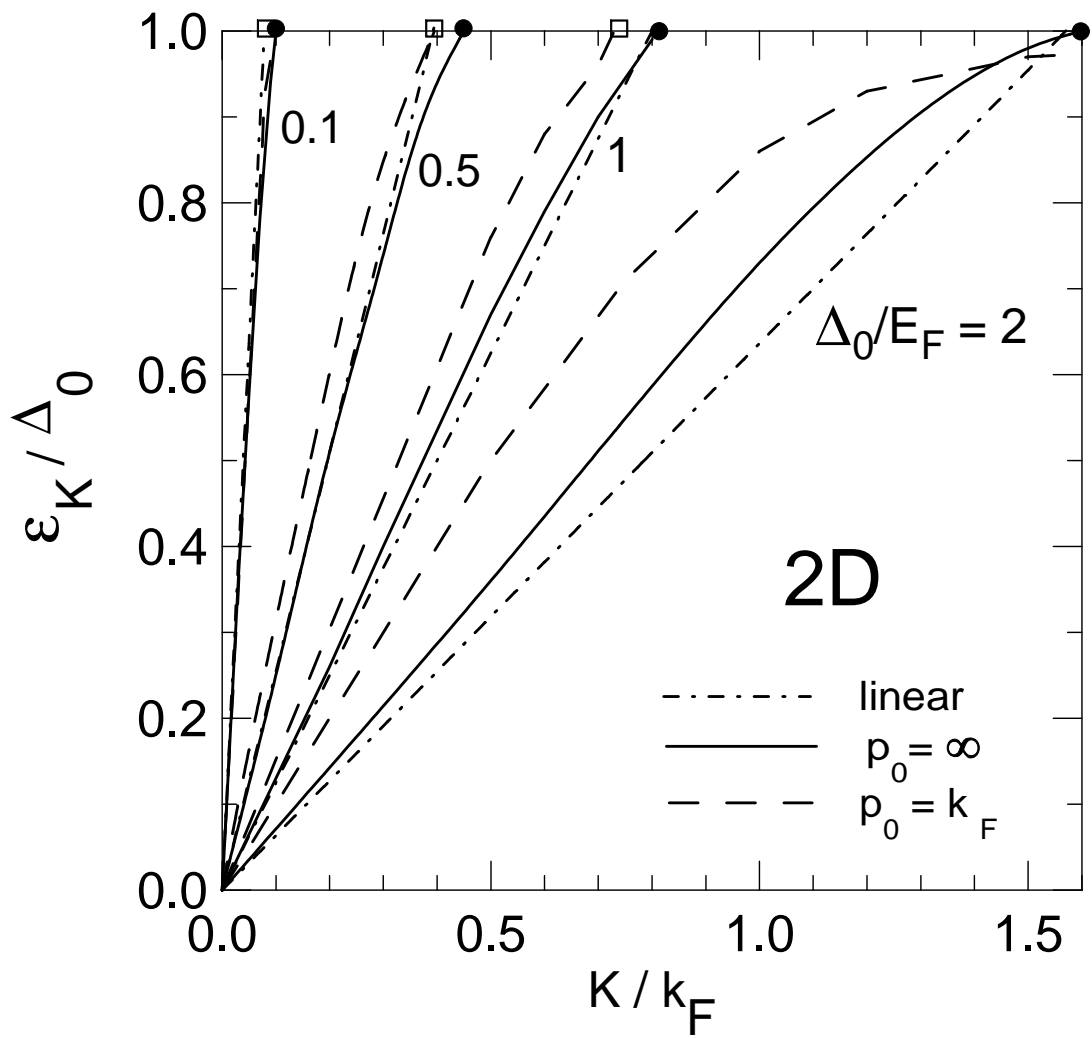
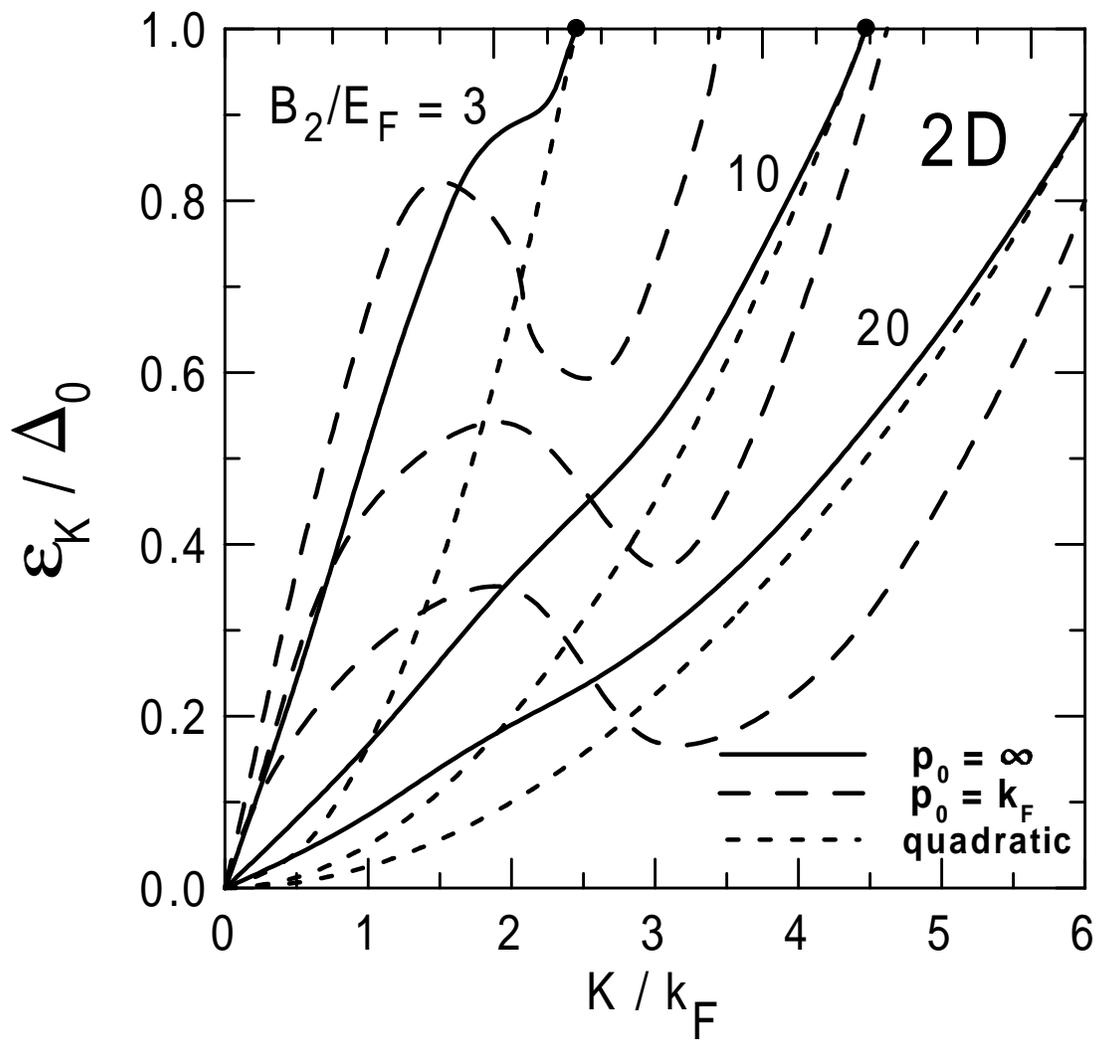
\end{document}